\documentclass[runningheads]{llncs}

\usepackage{graphicx}
\usepackage[utf8]{inputenc}
\usepackage{url}
\usepackage{hyperref}
\hypersetup{
    colorlinks,
    citecolor=blue,
    filecolor=blue,
    linkcolor=blue,
    urlcolor=blue
}

\usepackage{cite}
\usepackage{amsmath,amssymb,amsfonts}
\usepackage{algorithmic}
\usepackage{bbold}
\usepackage{bm}
\usepackage{bbm}
\usepackage[euler]{textgreek}
\usepackage{chngcntr}
\usepackage{listings}
\usepackage{fixltx2e}

\AtBeginDocument{
  \counterwithout{lstlisting}{chapter}%
}
\makeatletter
\renewcommand{\l@lstlisting}[2]{%
  \@dottedtocline{1}{0em}{1.5em}{\lstlistingname\ #1}{#2}%
}
\makeatother
\usepackage{wrapfig}

\DeclareMathOperator*{\bx}{{\mathbf{x}}}

\DeclareMathOperator*{\btheta}{{\bm{\theta}}}

\begin{document}

\title{Simulating Data Access Profiles of Computational Jobs in Data Grids}

\titlerunning{Simulating Data Access Profiles of Computational Jobs in Data Grids}

\author{Volodimir Begy\inst{1,3} \and
Joeri Hermans\inst{2} \and
Martin Barisits\inst{1} \and
Mario Lassnig\inst{1} \and
Erich Schikuta\inst{3}}

\authorrunning{V. Begy et al.}

\institute{CERN, Geneva, Switzerland\\
\email{volodimir.begy@cern.ch} \and
University of Li\`ege, Li\`ege, Belgium \and
University of Vienna, Vienna, Austria }

\maketitle

\begin{abstract}
The data access patterns of applications running in computing grids are changing due to the
recent proliferation of high speed local and wide area networks. The data-intensive jobs are no longer strictly
required to run at the computing sites, where the respective input data are located. Instead,
jobs may access the data employing arbitrary combinations of data-placement, stage-in and remote data access.
These data access profiles exhibit partially non-overlapping throughput bottlenecks. This fact can be
exploited in order to minimize the time jobs spend waiting for input data.
In this work we present a novel grid computing simulator, which puts a heavy emphasis on the various data access
profiles. The fundamental assumptions underlying our simulator are justified by empirical experiments performed
in the Worldwide LHC Computing Grid (WLCG) at CERN.
We demonstrate how to calibrate the simulator parameters in accordance with the true system using
posterior inference with likelihood-free Markov Chain Monte Carlo.
Thereafter, we validate the simulator's output with respect to an authentic production workload from WLCG,
demonstrating its remarkable accuracy.
\keywords{Grid Computing \and Data Access Patterns \and Network Modeling \and Discrete Event Simulation \and Bayesian Deep Learning \and
Likelihood-free Inference.}
\end{abstract}

\section{Introduction}

A large number of applications from different scientific fields relies on extensive computing resources.
These resources are provided by data centers, which are in turn aggregated to form
computing grids \cite{foster2003grid}.
Employing a wide range of heterogenous hardware, the participating computing sites work together to reach a
common goal in a coordinated manner. Grids store vast amounts of
scientific data, and numerous users run computational jobs to analyze these data in a highly distributed and
parallel fashion. For example, within
the World-Wide LHC Computing Grid (WLCG) more than 150 computing sites are employed by the ATLAS
experiment at CERN. WLCG stores more than 360 petabytes of ATLAS data, which is used for distributed analysis by more than 5000 users.

The grid resources are typically divided into three major classes: storage elements, worker
nodes and network.
Certain job types, e.g. Monte Carlo production in the
domain of High Energy Physics are data-intensive. These jobs are executed on data grids, which
posses higher storage capacities and network throughput.
The strict division of labour among classes of machines has led to the fact that the current best practice for data access in the
  grid is data-placement \cite{YUAN20101200}.
Given a job scheduled to run
at a specific data center, its execution may commence only after the completion of the following workflow.
First, the input data needs to be placed from the remote storage element to the local one. Secondly, the data
has to be staged-in from the local storage element into the worker node's scratch disk. Data-placement is handled by
distributed data management (DDM) systems.
For instance, the DDM system employed by the ATLAS experiment at CERN is
Rucio \cite{Garonne_2014}. An alternative approach is to
stream the input data from storage elements employing remote data access.
With the contemporary proliferation of high speed local and wide area networks remote data access is no longer
prohibitively expensive. In recent years numerous researchers have been examining its properties
in WLCG \cite{Elmsheuser_2015}.

In a grid computing setting it is challenging to design reproducible studies due to the highly dynamic nature
of the system. Furthermore, performance studies with high workloads interfere with the applications running
in production. A powerful tool to address these issues is a simulator, which is statistically tested
with respect to authentic logs from production workloads. In this work we propose a novel grid
computing simulator named GDAPS (Grid Data Access Profiles Simulator).
GDAPS is a discrete event simulator based on the established SimPy framework.
It supports modeling of the
3 common data access profiles in computing grids: data-placement, stage-in and remote data access.
Its source code is publicly available at \url{https://github.com/VolodimirBegy/GDAPS}.

The rest of the paper is organized as follows. In Section \ref{sota} we present related work
in the field of simulation of data-intensive systems. Thereafter, we demonstrate the results of
empirical experiments, which we have executed in the Worldwide LHC Computing Grid.
These results justify the fundamental
assumptions underlying the construction of GDAPS.
Section \ref{sec:archi} describes the architecture and the data transfer mechanism of the simulator.
Next, we evaluate the accuracy of GDAPS by simulating an authentic production workload.
Prior to the simulation the tool has to be calibrated in accordance with the true system.
We tune the simulator parameters responsible for the
parameterization of the system's latent processes.
For this purpose we rely on approximate Bayesian inference.
In particular, we perform posterior inference with likelihood-free Markov Chain Monte Carlo.

\section{Related Work}
\label{sota}

A large body of work in the literature addresses the construction of grid and cloud computing simulators
for respective performance studies.
However, none of these simulators provide the means to realistically model the behavior of
different data access profiles over LAN and WAN.
Most simulators model the network in overly naive and static ways and do not differentiate
between data transfer protocols. Furthermore, many authors either do not validate the accuracy of their simulators
against authentic traces or report poor results of such evaluations. Last but not least, many contributions are outdated with respect to
the recent advances in grid computing infrastructures. The purpose of the simulator
proposed in this work is to fill these gaps.
In the following paragraphs we describe the most notable grid and cloud computing simulators. We also
present performance studies implemented on top of these or custom simulators.

In \cite{buyya2002gridsim} Buyya et al. present GridSim. The tool models different grid components in a highly abstract
fashion, rendering it not suitable for specialized studies with a focus on networking.
In \cite{sulistio2008toolkit} Sulistio et al. extend GridSim with data grid functionality.
Concretely, the authors enable data querying, data replication and remote data access.
The authors in \cite{bell2003optorsim}
propose Optorsim in order to enable research on dynamic data replication. GangSim \cite{dumitrescu2005gangsim}
allows to implement scheduling and resource allocation policies. It also supports hierarchies of users.
However, the authors report large discrepancies between collected metrics in corresponding simulated
and real world experiments. Alea 2 \cite{klusavcek2010alea} provides a queue/plan based mechanism for evaluation of different scheduling algorithms.
The authors present simulations based on real traces, but do not report on their accuracy. GroudSim \cite{ostermann2010groudsim} is a framework for
modeling of scientific workflows, which are executed on combined
grid and cloud infrastructures. The toolkit improves the runtime performance of process-based analogues by employing
discrete-event simulation. ComBos \cite{alonso2017combos} simulates all components of BOINC, a middleware employed for volunteer and desktop grid
computing. A thorough treatment of the network infrastructures is left for the future work. Dobre et al. \cite{dobre2011monarc} present the Monarc (MOdels of Networked Analysis at
Regional Centers) simulation framework, which was developed at CERN for High Energy Physics
use-cases.
The authors only briefly discuss remote data access, and the demonstrated experiments
are based on numerous assumptions which are not tested.
CloudSim \cite{calheiros2011cloudsim} extends GridSim to provide cloud-related functionality. In particular,
it enables modeling of virtual machines and analysis of provisioning algorithms with respect to
Quality of Service parameters and Service Level Agreements. NetworkCloudSim \cite{garg2011networkcloudsim}
introduces different classes of switches (edge, aggregate and root) in order to model
the data centers' local area networks with a higher degree of precision.
EdgeNetworkCloudSim \cite{seufert2017edgenetworkcloudsim}
further introduces users, service chains and service request processing in the context of edge cloud computing.

The work in \cite{shah2011dynamic} presents multilevel hybrid scheduling algorithms based on a dynamic time quantum.
These scheduling techniques are evaluated on a dedicated simulator.
Ishii et al. \cite{ishii2011adaptive} evaluate the optimization of data access patterns in computing grids
 on top of a simulator.
The authors consider only data replication to local sites. Furthermore, the network dynamics are not
addressed in this paper. Camarasu-Pop et al. \cite{camarasu2016combining} analyze the effect of application
 makespan and checkpointing period
on Monte Carlo jobs in the
European Grid Infrastructure. The authors conduct production experiments in order to evaluate the proposed
analytical model and the simulator. Such extensive study allows to establish
robust and realistic results. In our work we ensure the quality of the simulator in an analogous manner.
The contribution in \cite{moghadam2016new} describes a 2-phase scheduling approach implemented
with Optorsim. In the first step a cluster is selected for job execution based on data access cost.
Secondly, a suitable worker from the cluster is chosen based on the task size and the node load.
The work in \cite{grace2016hgasa} presents HGASA, a hybrid heuristic for optimization
of data access patterns in computing grids based on a combination of genetic algorithms
and simulated annealing. The optimization is evaluated using GridSim.

\section{Empirical Analysis}
\label{analysis}

In this section we analyze certain aspects of the ATLAS data grid, which are relevant for the construction of a
realistic simulator. The ATLAS data grid is part of the Worldwide LHC Computing Grid. WLCG employs commodity hardware.
Thus, the findings from this section are universally generalizable.

Previous work \cite{Begy:2621616} has shown that
the throughput of remote data access can be modeled by the following linear regression:

\begin{equation}
T = 0 + a*S + b*ConTh + c*ConPr.
\label{eq:reg1}
\end{equation}

\noindent In the above equation \textit{T} is the transfer time of a file, \textit{S} is the file size, \textit{ConTh}
is the aggregated link traffic of concurrent threads within a given job and
\textit{ConPr} is the aggregated link traffic of concurrent processes within the investigated computational campaign.
The regression coefficients \textit{a}, \textit{b} and \textit{c} characterize the transfer throughput.
While a job may start multiple threads to stream different files remotely, when employing data-placement, each file
is transferred by an individual process. Thus, we formulate the hypothesis that the throughput of data-placement
can be modeled by the following linear regression:

\begin{equation}
T = 0 + a*S + b*ConPr.
\label{eq:reg2}
\end{equation}

\noindent To test this hypothesis, we select two random storage elements from WLCG, namely \textit{FZK-LCG2\_DATADISK} and
\textit{SLACXRD\_DATADISK}. We then query the Rucio Hadoop cluster for logged metrics on more than 27,000
\textit{gsiftp} file transfers between these two storage elements in the time window 02/05/2018 - 17/05/2018.
Through transformations we obtain the values of the variables \textit{T}, \textit{S} and \textit{ConPr}
for each file transfer.
Finally, we perform the linear regression, which results in the following fit:

\begin{equation}
T = 0.24045*S + 0.00044*ConPr.
\label{eq:reg3}
\end{equation}

\begin{figure}
\centering
\begin{minipage}{.495\textwidth}
  \centering
  \includegraphics[width=\textwidth]{dp_regr}
  \caption{Data-placement regression fit}
  \label{fig:grapher}
\end{minipage}\hfill
\begin{minipage}{.405\textwidth}
  \centering
  \includegraphics[width=\textwidth]{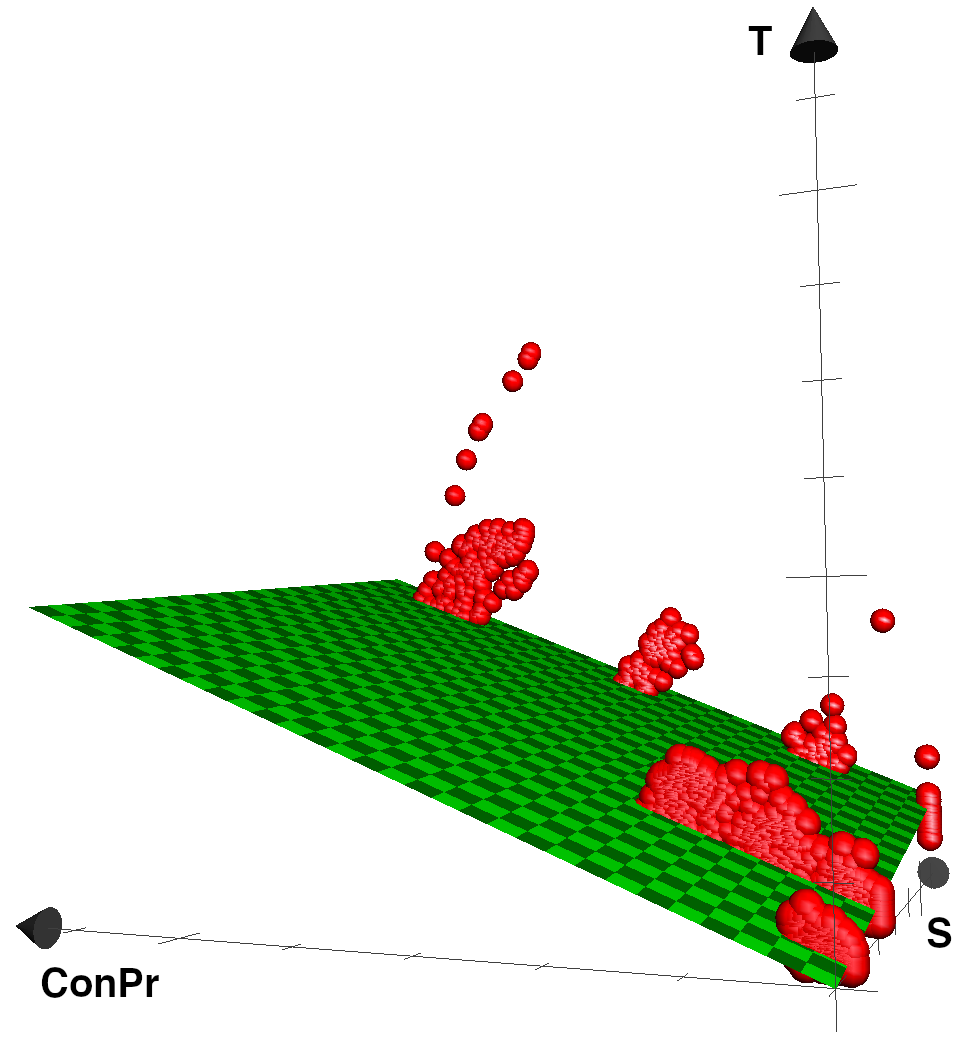}
  \caption{Stage-in regression fit}
  \label{fig:grapher2}
\end{minipage}
\end{figure}

\noindent  The fit has an F-statistic of 1.234e+05 on 2 degrees of freedom and 27021 residual degrees of freedom
and exhibits a p-value of \textless 2.2e-16. Thus, our hypothesis is confirmed. This proves that
the finding about the relationship among the variables generalizes well not only on data from our experiments,
but also on independently collected data.
 The regression fit is displayed together with the observations in Fig. \ref{fig:grapher}.

Next, we experimentally demonstrate that the linear regression from Eq. \ref{eq:reg2} also models
the throughput of stage-in transfers. In this experiment we repeatedly launch 1-12 concurrent jobs
on a worker node at the CERN data center (Switzerland) in the time window 08/08/2018 - 10/08/2018. Each job
launches a single process in order to stage-in files of different sizes (300MB - 3GB) using
the \textit{xrdcp} protocol. After data collection and transformation the resulting dataset has more than
2,000 observations and exhibits the following fit:

\begin{equation}
T = 0.036*S + 0.012*ConPr.
\label{eq:reg4}
\end{equation}

\noindent  The fit has an F-statistic of 8392 on 2 degrees of freedom and 2067 residual degrees of freedom and a p-value of
\textless 2.2e-16. It is displayed is Fig. \ref{fig:grapher2} along with the observations.

\begin{figure}
\centering
\includegraphics[width=.95\textwidth]{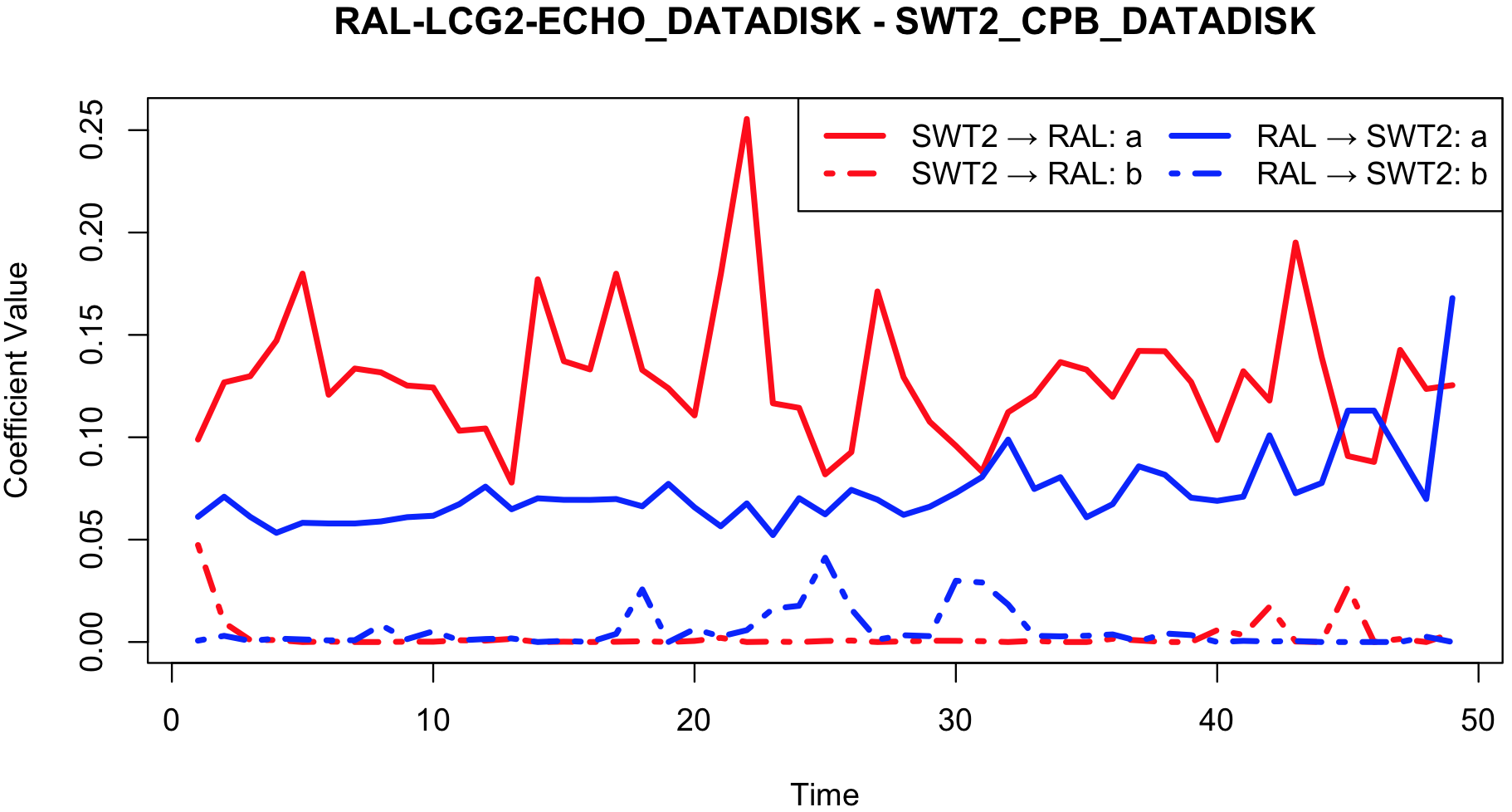}
\caption{Time series of regression coefficients \textit{a} and \textit{b} with hourly time resolution}
\label{fig:uni}
\end{figure}

In GDAPS the communication between two hosts occurs through a virtual link.
At the current implementation stage we exclusively consider data input. The
management of jobs' output data is neglected. Thus, a bi-directional link is
only possible between two storage elements.
In order to decide, whether a link between two storage elements needs to be modeled in an
uni- or bi-directional manner, we have performed the following analysis.
Two random ATLAS storage elements \textit{RAL-LCG2-ECHO\_DATADISK}
and \textit{SWT2\_CPB\_DATADISK} are selected. Then the Rucio Hadoop cluster is queried for logged metrics on all
\textit{gsiftp} transfers between these two storage elements in the time window 05/12/2018 23:00 - 08/12/2018 00:59.
The resulting dataset it partitioned hourly. The linear regression from Eq. \ref{eq:reg2} is applied to each
partition. The time series of regression coefficients \textit{a} and \textit{b} across all partitions are depicted
in Fig. \ref{fig:uni}.

\noindent Clearly, the coefficients do not exhibit a bi-directional throughput. We assume that this behavior is caused by the fact that
the traffic may take different paths over WAN when the source of data in a given host pair is switched.
Thus, GDAPS models virtual links in a uni-directional manner.

\section{Simulator Architecture}
\label{sec:archi}

 In this work we are focusing on data access patterns in computing grids.
 Thus, we primarily model and simulate components of data grids related to networking.
 The simulator is built based on assumptions, which are validated by
 empirical experiments and observations in the Worldwide LHC Computing Grid at CERN, as demonstrated
 in Section \ref{analysis}.

\begin{figure}
 \centering
 \includegraphics[width=\textwidth]{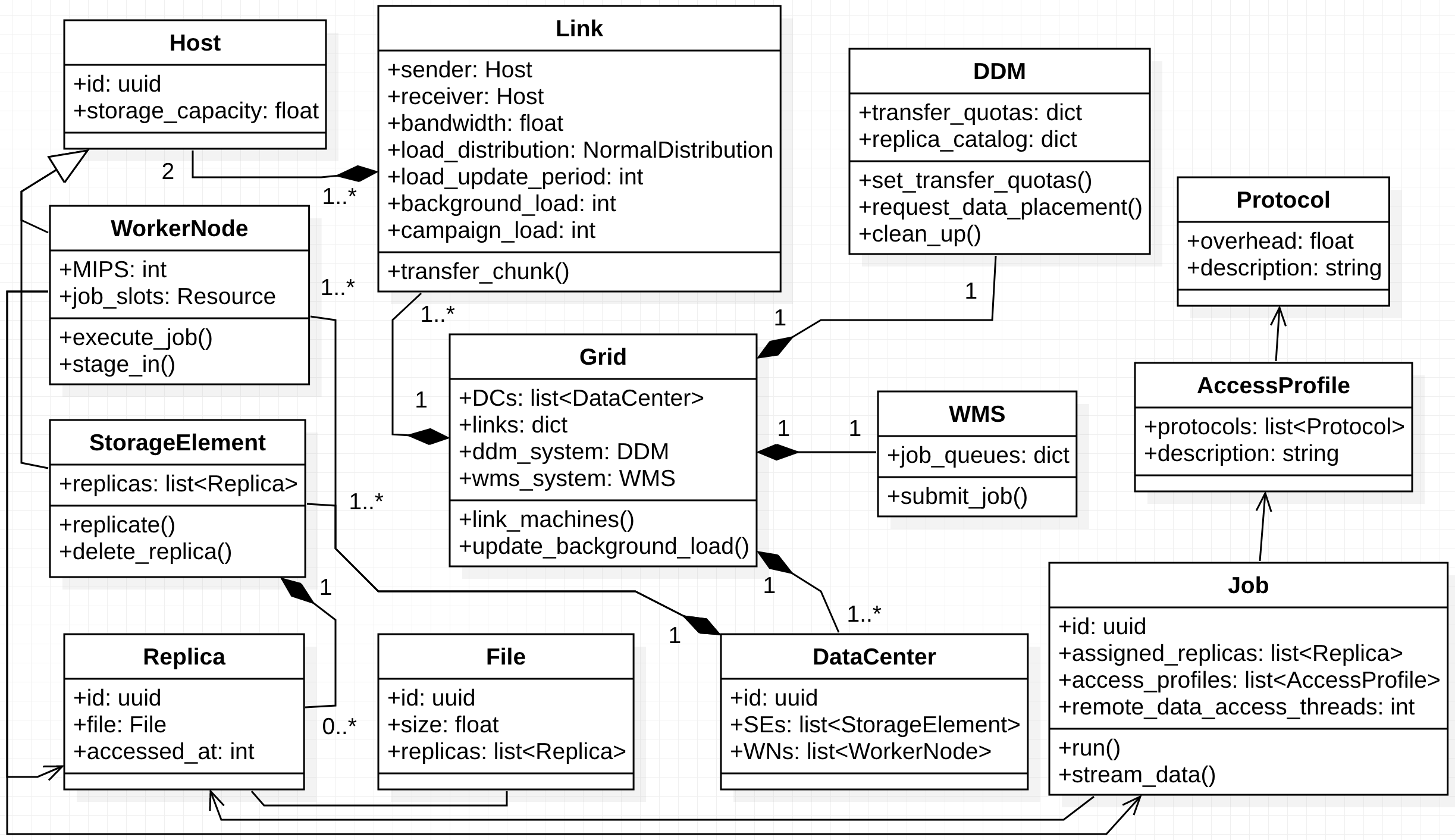}
 \caption{Class diagram of GDAPS}
 \label{fig:cd}
 \end{figure}

The architecture of GDAPS is presented in Fig. \ref{fig:cd} by a class diagram.
The central components of the simulator and their tasks are as follows.
Storage elements persist replicas of files for the long term.
Worker nodes execute computational jobs. Their performance is determined by the
\textit{million instructions per second} attribute. Worker nodes may also stage-in data from local storage elements
into their scratch disks.
Distributed Data Management System (DDM) is responsible for the monitoring
 of storage elements, enforcement of quotas on data-placement transfers and clean-up of outdated replicas.
Workload Management System (WMS) monitors worker nodes and submits jobs
 in accordance with the specified resource provisioning policies.
A virtual link has a pair of communicating hosts. Its fixed physical bandwidth
 is fairly allocated among all processes and threads.
The latent loads of a link are parameterized
 by a normal distribution and an update period.
Data centers are linked collections of storage elements and worker nodes. They are
 further aggregated to form the grid.
Replicas are realizations of files. They are persisted in storage elements and
 may be streamed remotely by jobs, staged-in by worker nodes or copied by the DDM.
An access profile is characterized by the employed data transfer protocols
 and the classes of the communicating hosts.
A computational job has a list of assigned replicas, which is synchronized with
 a list of respective access profiles. Within a single job numerous threads may
 concurrently stream input replicas.

A virtual connection between two hosts can be modeled with
various degrees of abstraction. Many factors affect the
data throughput of such connections, among others:
amount of concurrent data flows passing through a congested router;
configuration of routers (buffer settings, packet scheduling policies, \dots); characteristics of data transfer protocols
(stateful or stateless, congestion control, \dots); a single protocol may have differing implementations across various operating systems;
bandwidths, latencies and highly dynamic loads of all links across the communication path;
performance of hosts' network interface cards.
In practice, it is not possible to model communication links of a data grid with this extreme degree
of precision. The mentioned low-level metrics are typically not centrally accessible. Thus, following the
end-to-end arguments in system design \cite{saltzer1984end} we base our modeling
on metrics obtainable through experiments executed in the application layer.
A link in GDAPS transfers data in small chunks. Each chunk is transferred during a single simulation tick, which
abstracts a second. A link allocates its physical bandwidth equally to all concurrent processes.
The processes originate either from a simulated bag of jobs, or an unknown background load.
The background load is parameterized by a normal distribution and an update period.
It is able to encapsulate all latent processes affecting the link throughput.
Once per the update period a new value for the \textit{background load} attribute is sampled from
the respective distribution.
A computational process fairly divides the allocated bandwidth among its threads.
Finally, a small fraction of the chunk is neglected due to the coordination overhead of the data transfer protocol.
This mechanism is summarized by the following code snippet:
\begin{lstlisting}[breaklines=true,captionpos=b,label={lst:transf},language=Python,basicstyle=\ttfamily]
chunk = (link.bandwidth / (link.background_load + link.campaign_load)) / job.n_threads
chunk -= chunk*protocol.overhead
\end{lstlisting}

\section{Accuracy of the Simulator}

To demonstrate the accuracy of the proposed tool we simulate a production workload.
The structure of the authentic workload is as follows.
Various amounts of concurrent jobs (1-12) are assigned to a single worker node at the
CERN data center (Switzerland). Once per 15 minutes in the period of 28.04.2018 00:00 - 28.04.2018 06:15
the jobs initiate remote accesses to the storage element GRIF-LPNHE\_SCRATCHDISK at the GRIF-LPNHE data center (France).
The data transfer is realized by the WebDAV protocol.
At each step
the jobs launch various amounts of concurrent threads (up to 4).
The threads stream files of different sizes (300MB - 3GB). This allows us to sample a wide range of data for
the variables \textit{T}, \textit{S}, \textit{ConTh} and \textit{ConPr}.
Each launched file access is treated as an observation in the final dataset.
After such sampling and transformations we derive 106 authentic observations.
We then simulate the same workload in GDAPS, logging 106 simulated observations.
Thereafter, we apply the linear regression from Eq. \ref{eq:reg1} to both the datasets.
The true dataset is summarized by the following fit:

\begin{equation}
T = 0.02385*S + 0.04886*ConTh + 0.00117*ConPr.
\label{eq:reg5}
\end{equation}

\noindent The true fit has an F-statistic of 1.956e+04 on 3 degrees of freedom and 103 residual degrees of
freedom and a p-value of \textless 2.2e-16.
The regression coefficients of this fit characterize the data throughput of the production workload.
We assess the accuracy of the simulated coefficients based on the following error metric:

\begin{equation}
E(coef_\text{sim}) = abs(coef_\text{true} - coef_\text{sim}) / coef_\text{true}.
\label{eq:error}
\end{equation}

\noindent However, the following issue needs to be addressed before stochastic simulations can be executed.
The simulator parameters have to be calibrated with respect to the true system. In our study this
concerns the parameters of the data transfer mechanism.
While we could estimate the link bandwidth to be 10,000 Mbps, quantifying the protocol overhead or
determining the parameters ($\mu$ and \textsigma)
of the background load distribution are non-trivial tasks. These 3 parameters
form the simulator setting $\btheta$, which needs to be inferred:

\begin{equation}
\btheta =
\begin{pmatrix}
overhead\\
\mu\\
\sigma
\end{pmatrix}.
\label{eq:setting}
\end{equation}

\noindent The inference procedure will rely on true observations ${\bx}_\text{true}$, which are the regression coefficients
of the authentic fit:

\begin{equation}
{\bx}_\text{true} =
\begin{pmatrix}
a_\text{true}\\
b_\text{true}\\
c_\text{true}
\end{pmatrix}
=
\begin{pmatrix}
0.02385\\
0.04886\\
0.00117
\end{pmatrix}.
\label{eq:xtrue}
\end{equation}

\noindent The Bayes' rule allows one to calculate the posterior distribution $p(\btheta \vert \bx)$ analytically.
However, in our study neither the likelihood $p(\bx \vert \btheta)$, nor the marginal $p(\bx)$
are tractable.
A further difficulty lies in the fact
that our generative model, which produces simulated observations ${\bx}_\text{sim}$ is a non-differentiable simulator.
To estimate the posterior in this setting we employ likelihood-free Markov Chain Monte Carlo
with approximate likelihood ratios \cite{2019arXiv190304057H, hypothesis}. In this method,
a parameterized classifier is trained to distinguish samples from the marginal
$p(\bx)$ and the likelihood $p(\bx \vert \btheta)$.
The classifier's output is then used to construct likelihood ratios $r(\bx, \btheta_{t}, \btheta')$
across different parameter settings, which represent the states of the Markov Chain.
The likelihood ratios are in turn employed by MCMC to either accept a proposed state ${\btheta}'$ or
once again sample the current state ${\btheta}_{t}$.
Once the sampling is completed, the histograms over the chain's states approximate the posterior
density $p(\btheta \vert \bx)$.

We realize the parameterized classifier by a deep
neural network with 4 hidden layers, 128 hidden units and SELU nonlinearities.
For each simulator parameter we have assumed a uniform prior distribution with the following bounds:
\begin{itemize}
\item WebDAV overhead: (0, 0.1)
\item $\mu$, the mean of the background load distribution: (0, 100)
\item \textsigma, the standard deviation of the background load distribution: (0, 100)
\end{itemize}

\begin{figure}
\centering
\begin{minipage}{.49\textwidth}
  \centering
  \includegraphics[width=\textwidth]{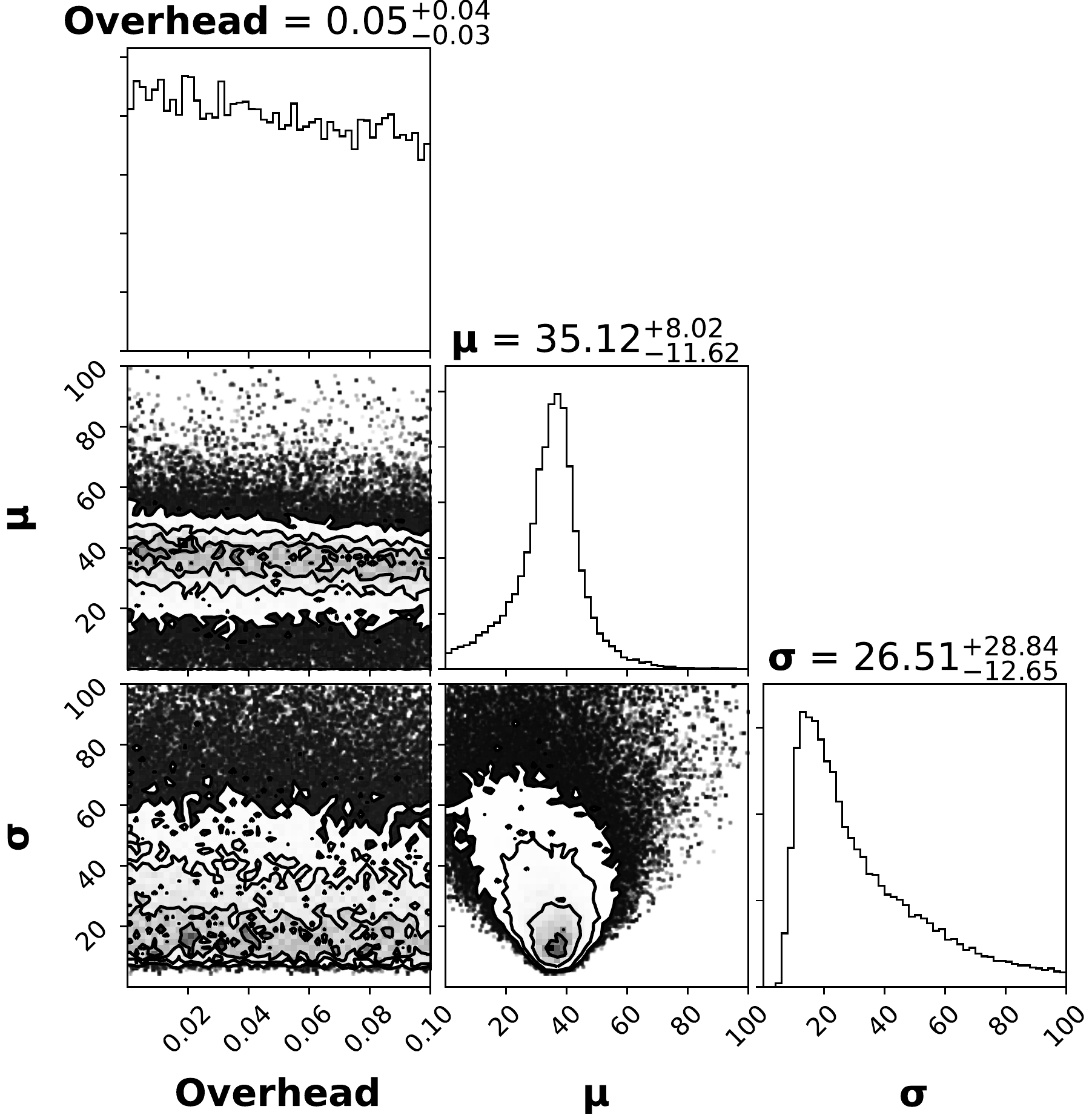}
  \caption{Approximate posterior density over the simulator setting $\btheta$}
  \label{fig:posterior}
\end{minipage}
\hfill
\begin{minipage}{.484\textwidth}
  \centering
  \includegraphics[width=\textwidth]{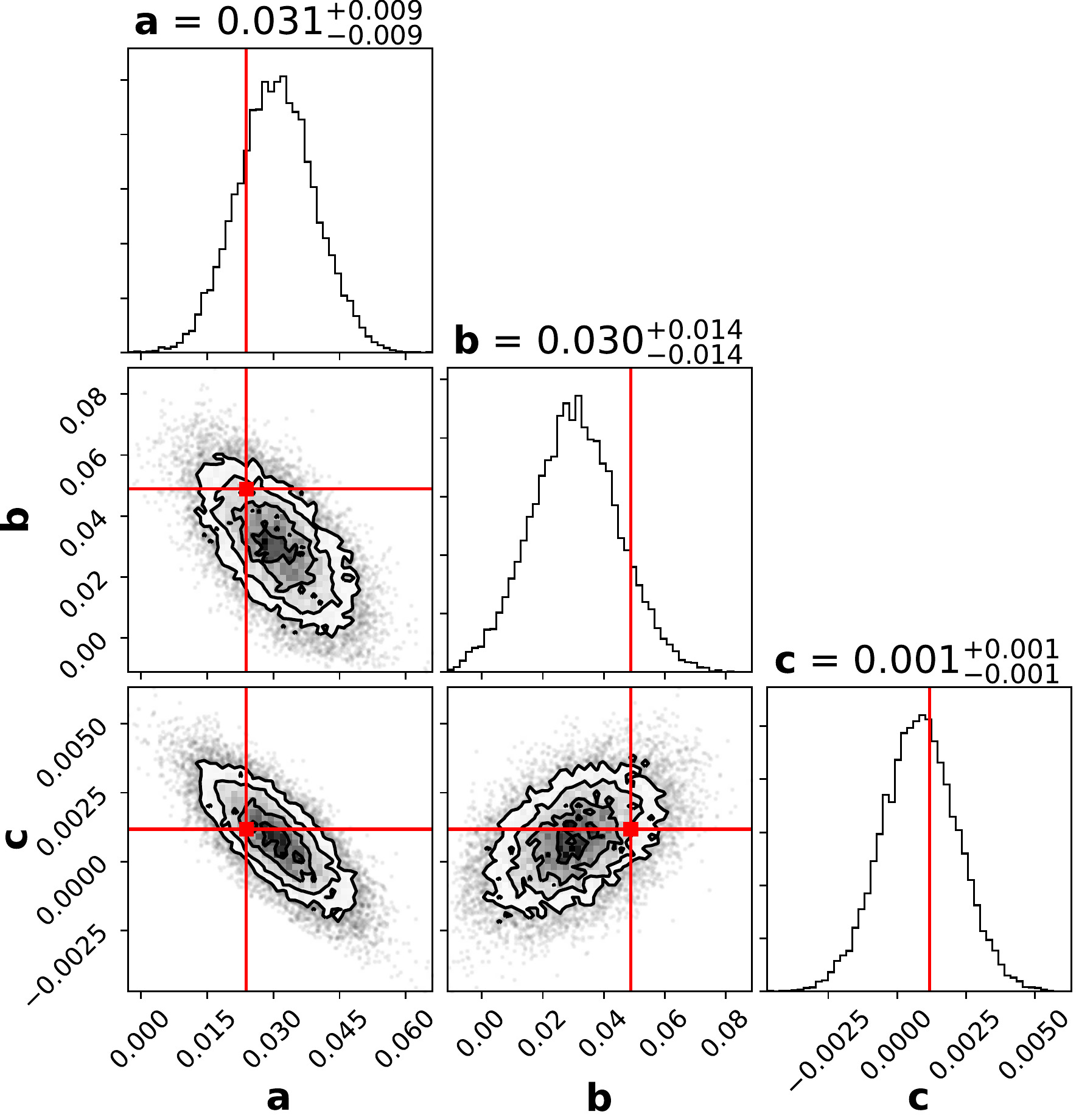}
  \caption{Coefficients simulated under the optimal parameter setting ${\btheta}^*$}
  \label{fig:sim_coefs}
\end{minipage}
\end{figure}

\noindent Given these priors we have pre-simulated more than 12.7 million of ($\btheta, {\bx}_\text{sim}$)-tuples, which
form the training set. The dataset is projected onto the interval (0,1)
to stabilize the training. The net is trained for 263 epochs using the ADAM optimization
algorithm with a learning rate of 0.0001.
Once the classifier is trained, we start the posterior MCMC sampling in the middle of the prior bounds.
Firstly, we sample 100,000 burn-in states in order to reach a stable region in the parameter space.
Thereafter, we collect 1,000,000 actual MCMC samples.
The histograms of the resulting Markov Chain approximate the multivariate posterior density. These
histograms are presented along with their covariances by a cornerplot in Fig. \ref{fig:posterior}.
The 0.5 quantile is reported above each histogram.
While the density of the overhead parameter is almost uniform,
clear modes are identified for $\mu$ and $\sigma$.
Given this approximate posterior, we pick the optimal simulator setting ${\btheta}^*$ which maximizes the
density along each axis:

\begin{equation}
{\btheta}^* =
\begin{pmatrix}
overhead^*\\
\mu^*\\
\sigma^*
\end{pmatrix}
=
\begin{pmatrix}
0.02\\
36.9\\
14.4
\end{pmatrix}.
\label{eq:thetaopt}
\end{equation}

\noindent Employing the optimal parameter setting ${\btheta}^*$ we execute 16,000 stochastic simulations
of the production workload.
At the end of each simulation we regress the produced dataset the same way as the authentic dataset.
The simulated coefficients are accumulated to form samples.
The histograms of these samples are shown along with their covariances in Fig. \ref{fig:sim_coefs}.
The 0.5 quantile is indicated above each histogram.
The values of the true coefficients are superimposed in red.

\noindent Since the distributions of the simulated coefficients
recover the true coefficients, we conclude that the likelihood-free
MCMC has succeeded in calibrating the simulator parameters. Based on our experience, the
 naive picking of random parameter settings is unable to perform this task.
Thus, it is evident that GDAPS executes highly realistic simulations.
A number of concrete tuples of simulated coefficients are
presented in Table \ref{tab1} along with the respective errors. The errors
are calculated using the metric from Eq. \ref{eq:error}.

\begin{table}
\caption{Example simulated coefficients}
\begin{center}
\begin{tabular}{ | c | c| c | c | c | c | c | }
\hline
\textbf{a\textsubscript{sim}}  & \textbf{E(a\textsubscript{sim})}  & \textbf{b\textsubscript{sim}} & \textbf{E(b\textsubscript{sim})} & \textbf{c\textsubscript{sim}} & \textbf{E(c\textsubscript{sim})} & \textbf{ \( \sum \) E} \\
\hline
0.02352 & 1.4\% & 0.049 & 0.3\%  & 0.00114 & 3.3\% & 5\%  \\
\hline
0.02427 & 1.7\% & 0.05038 & 3.1\%  & 0.00118 & 0.4\% & 5.2\%  \\
\hline
0.02408 & 0.9\% & 0.05006 & 2.5\%  & 0.00121 & 3.1\% & 6.5\%  \\
\hline
0.02477 & 3.8\% & 0.04810 & 1.6\%  & 0.001206 & 2.8\% & 8.2\%  \\
\hline
0.02458 & 3\% & 0.04668 & 4.5\%  & 0.00116 & 0.7\% & 8.2\%  \\
\hline
0.02467 & 3.4\% & 0.04693 & 4\%  & 0.001185 & 1\% & 8.4\%  \\
\hline
0.02538 & 6.4\% & 0.04854 & 0.7\%  & 0.00115 & 2\% & 9.1\%  \\
\hline
0.02298 & 3.6\% & 0.05121 & 4.8\%  & 0.00118 & 0.8\% & 9.2\%  \\
\hline
\end{tabular}
\label{tab1}
\end{center}
\end{table}

\section{Conclusions and Future Work}
In this paper we have presented a novel grid computing simulator.
GDAPS allows one to model and simulate data access profiles of computing jobs in data-intensive systems.
The fundamental assumptions underlying the construction of our simulator are justified by empirical experiments
performed in the Worldwide LHC Computing Grid.

To study the accuracy of the simulations produced by our tool we have executed an authentic production workload from
WLCG.
Prior to the simulations we have successfully calibrated the simulator parameters with respect to the true system.
The calibration relied on approximate Bayesian inference with likelihood-free Markov Chain Monte Carlo.
It allowed us to obtain the approximate posterior density over a set of simulator parameters, which affect the data
transfer mechanism.
Given the optimal parameter setting ${\btheta}^*$, the distributions of the simulated regression coefficients
were able to recover the true coefficients.

The future work will confront the thorough modeling of the ATLAS data grid in GDAPS.
Thereafter, we will perform evolutionary optimization of data access patterns in
bags of jobs with the objective to minimize the joint data transfer time. This constitutes a
constrained optimization problem. The fitness of proposed solutions will be evaluated on top
of GDAPS, since we can rely on its accuracy.
Lastly, we will extend the simulator with further functionality, such as management
of jobs' output data, support for more resource provisioning algorithms and virtual organizations.

\bibliographystyle{splncs04}
\bibliography{refs}

\end{document}